\documentclass[12pt]{article}     
\usepackage{amsmath,epsfig,amssymb,eufrak}

\input eqalign.sty     

\date{}

\newcommand{\be}{\begin{equation}}
\newcommand{\ee}{\end{equation}}
\newcommand{\bea}{\begin{eqnarray}}
\newcommand{\eea}{\end{eqnarray}}

\newcommand{\dds}{\stackrel{\leftrightarrow}{D}}

\begin{document}

\begin{titlepage}

\title{
  {\vspace{-0cm} \normalsize
  \hfill \parbox{40mm}{CERN-TH/2000-259}\\
  \hfill \parbox{40mm}{NIC/DESY-00-001}}\\[30mm]
  Lattice hadron matrix elements with the Schr\"odinger functional:
the case of the first moment of non-singlet quark density\\ }
  \author{M.\ Guagnelli$^{a}$, K.\ Jansen$^{b}$ and R.\ Petronzio$^{a}$ \\
  {\small $^a$ Dipartimento di Fisica, Universit\`a di Roma 
	{\em Tor Vergata}}\\ 
  {\small and INFN, Sezione di Roma II,} \\
  {\small Via della Ricerca Scientifica 1, 00133 Rome, Italy} \\
  {\small $^b$ CERN, Theory Division, CH-1211 Geneva 23, Switzerland}\\
}

\maketitle

\begin{abstract}

We present the results of a non-perturbative determination of the pion
matrix element of the twist-2 operator corresponding to the average
momentum of non-singlet quark densities. The calculation is made within
the Schr\"odinger functional scheme. We report the results of simulations
done with the standard Wilson action and with the non-perturbatively
improved clover action and we show that their ratio correctly extrapolates,
in the continuum limit, to a value compatible with the residual correction
factor expected from perturbation theory.
\vspace{0.8 cm}
\noindent
\end{abstract}

\vfill
\begin{flushleft}
\begin{minipage}[t]{5. cm}
  {CERN-TH/2000-259}\\
  August 2000
\end{minipage}
\end{flushleft}
\end{titlepage}

\pagebreak

The calculation of hadronic matrix elements of Wilson operators entering
the light cone expansion of two electroweak currents, directly connected to
the moments of parton density distributions, requires non-perturbative
tools. Lattice estimates have been produced so far for only the first few
such moments. The results for the
second moment of non-singlet quark densities, corresponding to the average momentum,
are typically higher than the experimental values.  However, these results have
been obtained at fixed lattice spacing only and using, as
renormalization factor for the bare operator, the one obtained from a
perturbative calculation. It is the aim of our alternative approach
to remove these approximations.

The scope of the present paper is to provide values of the bare matrix element at
different lattice spacing. After  
multiplication with the corresponding non-perturbative 
renormalization factor,
these values can be used for an estimate of the continuum
limit of the renormalized matrix element. 
In a series of papers, other essential
ingredients needed for such an estimate were obtained. 
In particular, the Schr\"odinger
functional (SF) renormalization scheme was discussed in
\cite{ref:perturbative}, the non-perturbative step scaling function describing the
scale evolution of the continuum
renormalization factor in \cite{ref:non-pert}, the universality of the
continuum limit of the step scaling function in \cite{ref:universal} and the
definition of a ``renormalization group invariant'' step scaling function 
in \cite{ref:invariant}.  The pion matrix element has already been
calculated for the Wilson action and with standard periodic boundary
conditions \cite{ref:schierholz}: here we want to show how the SF can be used to reliably 
calculate not only the renormalization factor but also the physical matrix element itself.



The use of the SF for extracting hadron correlation functions has been
initiated with the calculation of pseudoscalar and vector masses and
decay constants by the ALPHA collaboration~\cite{ref:alpha_matrix}.  The
calculation with the SF of a two-quark matrix element has never been
attempted before and, with respect to a traditional method, presents
two advantages.  The first is the possibility of
constructing ``smeared'' and gauge-invariant states as particle sources at
the boundary (for the pion in this case). 
The second advantage is that 
the choice of SF boundary
conditions allows insertion of the operator at $T/2$, where $T$ is the total
time extent of the lattice, to be compared with the maximum value $T/4$,
reachable with ordinary periodic boundary conditions. If the operator of interest
is inserted at a fixed physical distance, this amounts to needing a lattice
for the SF
that is a factor of 2 smaller in the time direction than the one using periodic boundary
conditions. 

In our previous work we have evaluated a non-perturbative renormalization factor of 
the following operator:

\begin{equation}
{\cal O}_{12}(x) = \frac{1}{4}\bar\psi(x) \gamma_{\{1} \dds_{2\}} \frac{\tau^3}{2}\psi(x)\; ,
\label{eq:operator}
\end{equation}

\noindent where $\dds_{\mu}$ is the covariant derivative and the bracket
around indices means symmetrization.
The calculation of the matrix element of this operator requires a non-zero pion momentum. 
In practice, this leads to a very noisy 
signal for large time separation.

This has been observed also with periodic boundary conditions in ref.
\cite{ref:schierholz}, but is even more crucial in our case, where we can reach
a larger distance in time to project on the lightest state.  We therefore decided to follow
ref.~\cite{ref:schierholz} and compute numerically on the lattice the
desired matrix element between pion states with a different lattice
representation of the twist-2 operator than the one defined in
eq.~(\ref{eq:operator}). This amounts to taking:

\begin{equation}
{\cal O}_{00}(x) = \frac{1}{4}\bar\psi(x)\left[ 
\gamma_0 \dds_0 - \frac{1}{3}\sum_{k=1}^3 \gamma_k \dds_k\right]\frac{\tau^3}{2}\psi(x)\; .
\label{eq:operator_44}
\end{equation}

The advantage of this operator is that it can be computed at vanishing
external momentum and is hence expected to show a much better signal to
noise behaviour than ${\cal O}_{12}$ and not to be contaminated too strongly by
lattice artefacts.  Indeed, as in \cite{ref:schierholz}, we find also within
our SF calculation, the correlation function of ${\cal O}_{00}$ to be much
less noisy than ${\cal O}_{12}$. In addition, we want to remark that we find with comparable statistics,
the error of the matrixelement of the operator ${\cal O}_{00}$ to be considerably smaller
when using SF than for periodic boundary conditions as used in \cite{ref:schierholz}.


\begin{figure}
\vspace{0.0cm}
\begin{center}
\psfig{file=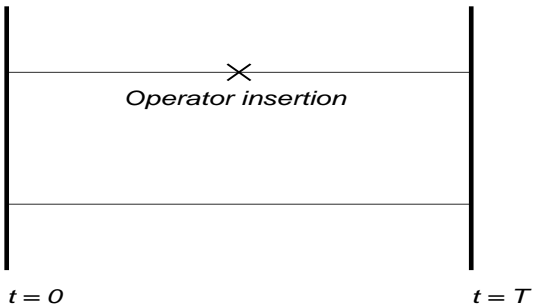, width=7cm,height=4cm}
\end{center}
\caption{ \label{fig:matrix_diagram} 
The correlation function
}
\end{figure}

The operator is inserted between two SF states defined at the two time
boundaries, as depicted in fig.~\ref{fig:matrix_diagram}.
We start by defining the boundary operators at the time boundaries $0$ and
$T$:
\begin{equation}
{\cal O}_{0} = \frac{a^6}{L^3}\sum_{\bf{y},\bf{z}}
\bar\zeta({\bf{y}})\gamma_5 \frac{\tau^3}{2}\zeta({\bf{z}}),\quad\quad\quad
{\cal O}_{T} = \frac{a^6}{L^3}\sum_{\bf{y},\bf{z}}
\bar\zeta'({\bf{y}})\gamma_5 \frac{\tau^3}{2}\zeta'({\bf{z}})
\end{equation}
and then the correlation function (of mass dimension equal to 1):
\begin{equation}
f_{\rm M}(x_0) = a^3\sum_{\bf{x}} 
\langle {\cal O}_{0} {\cal O}_{00}(x) {\cal O}_{T} \rangle\; .
\end{equation}
We also consider, for purposes of normalization, the correlation function
\begin{equation}
f_{1} = -\langle {\cal O}_{0} {\cal O}_{T} \rangle\; .
\end{equation}

Closely following the discussion in ref.~\cite{ref:alpha_matrix}, we easily see that
for a time extent large enough, the matrix element is expected to be taken
among the lightest particles coupled to the surface states, i.e. between
pions, in our case.  The general formulae, including the first excited state,
are:
\begin{eqnarray}
f_1 &\simeq& \rho^2e^{-m_\pi T}, \nonumber \\
f_{\rm M}(x_0) &\simeq& \rho^2e^{-m_\pi T}\langle \pi | {\cal O}_{00} | \pi \rangle 
\left\{ 1 + \rho'e^{-\Delta x_0} + \rho'' e^{-\Delta (T-x_0)}\right \}\; .
\end{eqnarray}
The actual matrix element is obtained from $f_{\rm M}(x_0)$ only after a suitable normalization
by $f_1$, which takes out the effects (wave-function contribution) of the boundary
quark fields. 
Assuming that there is a plateau region where 
$f_M(x_0) / f_1 = \mathrm{const} \equiv \langle \pi | {\cal O}_{00} | \pi \rangle$, 
and in which the first excited state
gives essentially no contributions, we obtain the physical matrix element by
a suitable normalization (see \cite{ref:schierholz}):

\begin{equation}
\langle x \rangle \equiv \frac{2\kappa}{m_\pi}\langle \pi | {\cal O}_{00} | \pi \rangle\; ,
\label{eq:ratio}
\end{equation}
with $\kappa$ the hopping parameter appearing in the fermion action.
The value of the pion mass
is obtained, following ref.~\cite{ref:alpha_matrix}, from the time dependence
of the pseudoscalar ($f_{\rm P}$) and/or axial-vector ($f_{\rm A}$) correlation functions.

\begin{figure}
\vspace{0.0cm}
\begin{center}
\psfig{file=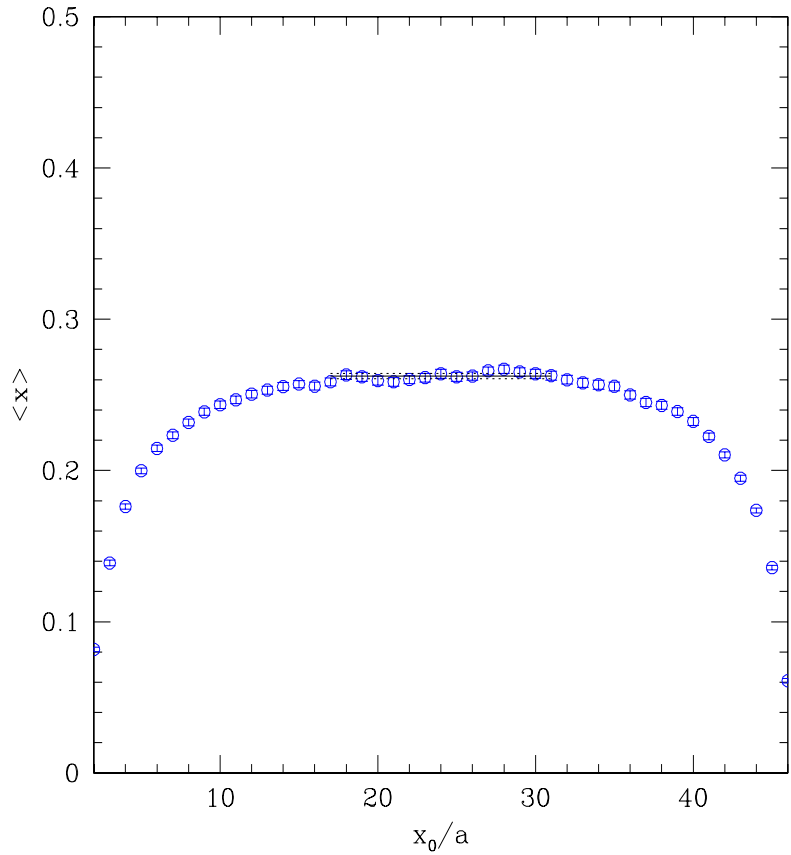, width=11cm,height=12cm}
\end{center}
\caption{ \label{fig:plateau} Example for the plateau behaviour of the
correlation function $2\kappa/m_\pi\cdot f_{\rm M}(x_0)/f_1$ 
to extract the matrix element, taken at $\beta=6.2$,
$\kappa=0.1346$ on a $24^3\cdot 48$ lattice (clover action).  }
\end{figure}

The set-up for our numerical simulation, performed in the quenched approximation,
was to choose lattices of physical
size $L^3\cdot T$, with $T$ taken to be $T\approx 3\;{\rm fm}$ in all
calculations.  We performed two sets of simulations, one using the Wilson
action and a second, using the non-perturbatively improved clover action.
The correlation
function projects on a single pion state at a distance in time
corresponding to about 1 fm. We then selected a time interval of 1 fm
again to extract the matrix element. An example for the plateau behaviour
of our correlation function and a fit to a plateau value with a distance of
about 1 fm is shown in fig.~\ref{fig:plateau}.


\begin{figure}
\vspace{0.0cm}
\begin{center}
\psfig{file=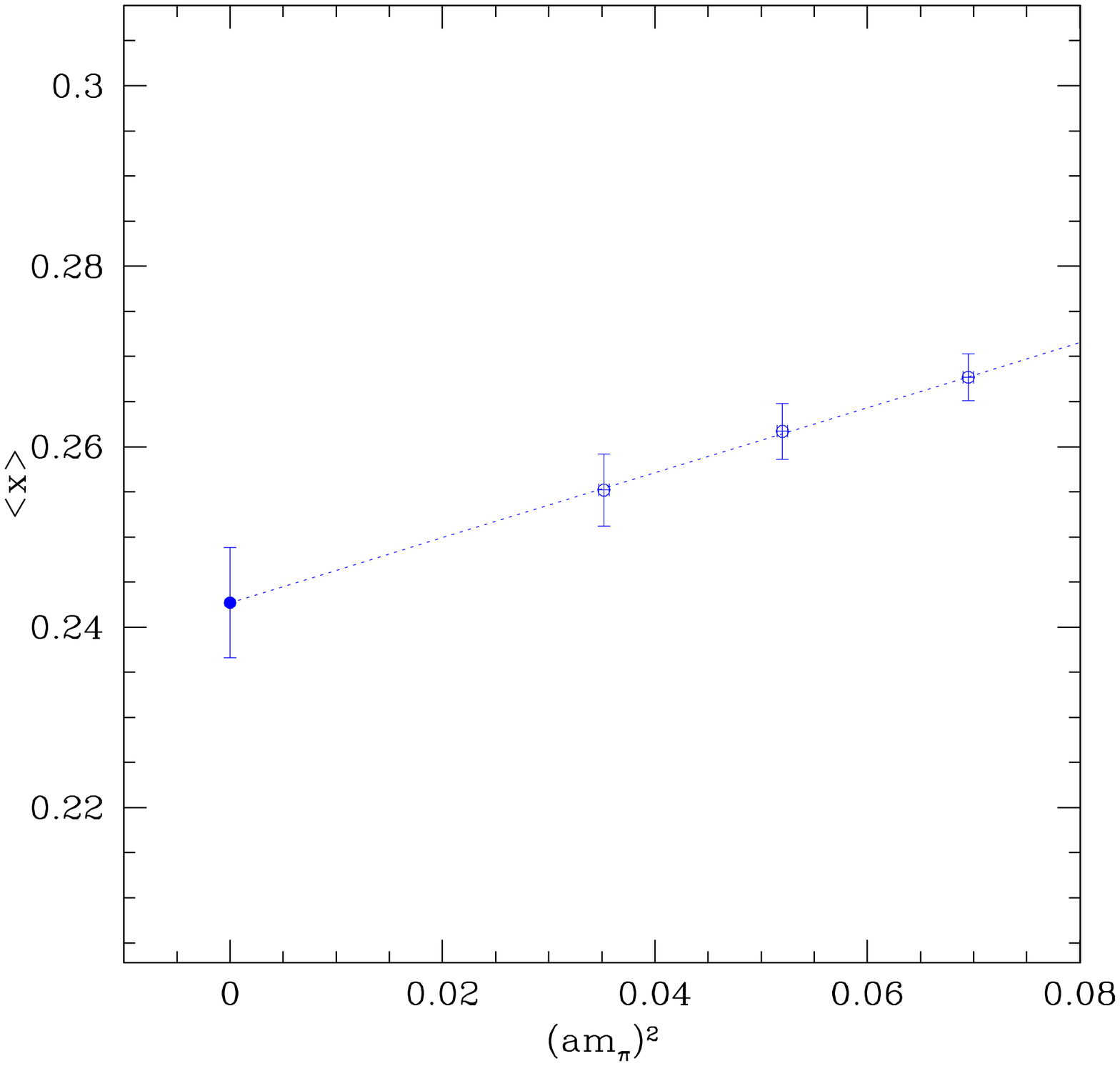, width=11cm,height=12cm}
\end{center}
\caption{ \label{fig:chiral}
Example for the chiral extrapolation of the matrix element at 
$\beta=6.3$ using the clover action. 
}
\end{figure}

At each of our four values of $\beta=6/g_0^2$, the calculation of the
matrix element is performed at three values of the quark masses, using a multiple mass solver.
Table~\ref{tab:allres} contains all our results for the matrix
element $\langle x\rangle$ at various values of the quark mass
and of $\beta$, for both the Wilson and for the $\mathrm{O}(a)$-improved clover
action.  In fig.~\ref{fig:chiral} we show one example for the chiral
extrapolation of the matrix element.


\begin{figure}
\vspace{0.0cm}
\begin{center}
\psfig{file=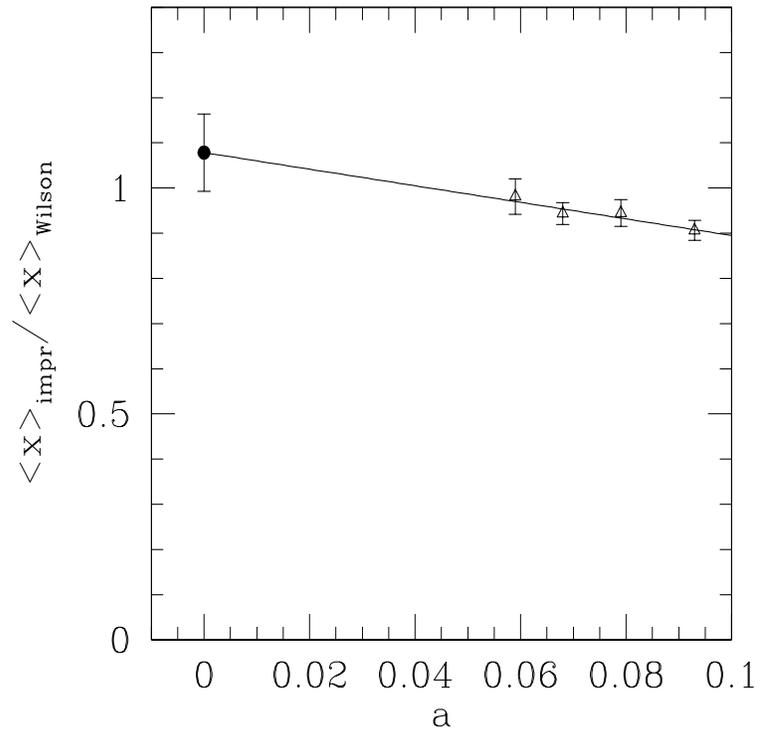, width=11cm,height=12cm}
\end{center}
\caption{ \label{fig:combined_fit}
The continuum extrapolation of the ratio of the bare lattice matrix
elements computed with the clover--improved and Wilson fermion action.
}
\end{figure}

In 
table~\ref{tab:mainres} we give the data for the matrix elements extrapolated to
the chiral limit. They can now be used to test the continuum extrapolation, by
building the ratio of the matrix elements calculated with the two different lattice
actions we have used in our simulation. 
In fig.~\ref{fig:combined_fit} we show these ratios as a function of the
lattice spacing. 
The data are extrapolated to the continuum as a linear function
of the lattice spacing. Given the limited range in the
bare coupling where the data were collected, the extrapolation procedure
only extrapolates lattice artefacts that vanish like a power of the lattice
spacing. Logarithmically dependent corrections, {\em i.e.} terms
proportional to the bare coupling, cannot be taken into account in this way.
Then, the result of the
continuum extrapolation of the ratio is not expected to be one, but the
value of a correction factor evaluated at a
bare coupling value of order $1$, corresponding to the region in $\beta$
where the extrapolation was
made.  The correction factor from a perturbative calculation at $g^2= 1$ is
$1.04$~\cite{ref:schpert}.  
Figure~\ref{fig:combined_fit} shows that our data are very well
compatible with such a value and gives us confidence in the uniqueness of the
continuum limit.

\begin{figure}
\vspace{0.0cm}
\begin{center}
\psfig{file=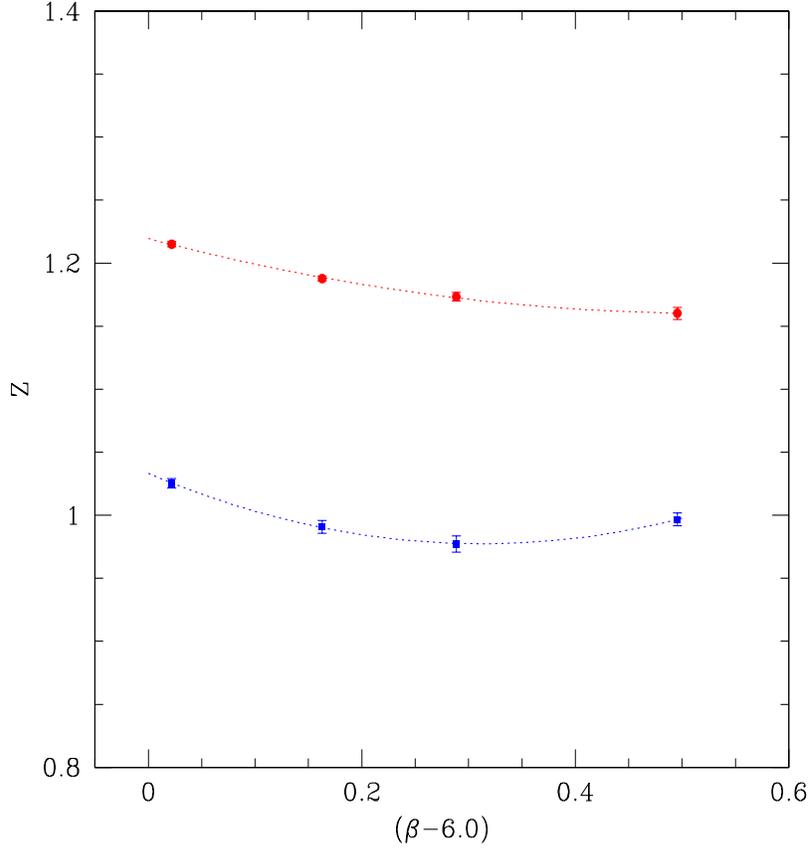, width=11cm,height=12cm}
\end{center}
\caption{ \label{fig:zeta}
Non-perturbative evaluation of $Z_{12}$ at a fixed inverse scale $L = 1.436\;r_0$.
The upper curve is for clover-improved fermions, 
the lower one for Wilson fermions. The numbers quoted in
table~\ref{tab:mainres} follow from the following interpolations (shown as dotted lines), 
which
we take as a definition of the two renormalization constants in this
range of $\beta$ values:
\newline
$Z_{12}^{\rm C} = 1.2196 - 0.2244(\beta-6) + 0.2117(\beta-6)^2,$
\newline 
$Z_{12}^{\rm W} = 1.0331 - 0.3570(\beta-6) + 0.5714(\beta-6)^2$.} 
\end{figure}

We have also computed the renormalization factor $Z_{12}$ of the operator 
of eq.~(\ref{eq:operator}) at the smallest scale $\mu_{\mathrm{min}}$ we could reach in our 
previous work for the scale evolution, $\mu_{\mathrm{min}}^{-1}=1.436r_0$
with $r_0\approx0.5\;\mathrm{fm}$. 
The values of $Z_{12}$ for our two lattice actions are shown in 
fig.~\ref{fig:zeta} as a function of $\beta$. In the figure the physical scale
$\mu_{\mathrm{min}}$ is kept fixed when varying $\beta$ by suitably choosing the lattice
size. 
In order to obtain the renormalization constant at exactly the values
of $\beta$ where the matrix element is computed, we performed an interpolation
which is also given in the caption of fig.~\ref{fig:zeta}. We give the interpolated
values of $Z_{12}$ in table~\ref{tab:mainres}.

If we would now use $Z_{12}$ to renormalize the operator ${\cal O}_{00}$,
the extrapolation to the continuum limit of the renormalized 
pion matrix element of the operator ${\cal O}_{00}$
would acquire a similar correction factor, 
as in the case of the ratio of the bare matrix element as discussed above.
This is why we do not proceed to extract directly from our matrix element
its renormalized value in the continuum limit and we postpone to a
forthcoming paper the non-perturbative evaluation of such a correction
factor, which anyway is expected, from perturbation theory, to be of the
order of a few per cent.

We have demonstrated in this paper that 
the SF calculation of hadron matrix elements is feasible and could be
applied to other interesting cases, such as the operators arising from
effective weak hamiltonians.


\begin{table}[htbp]
\begin{center}
\leavevmode
\begin{tabular}[]{|c|c|c|c|c|c|}
\hline
$\beta$ & $\kappa$ & Action & fit interval & $m_{\pi}$    & $\langle x\rangle$ \\
\hline\hline
        & $0.153$  &        &              & $0.4183(14)$ & $0.3096(17)$ \\
$6.0$   & $0.154$  & W      & $11-21$      & $0.3594(15)$ & $0.3005(22)$ \\
        & $0.155$  &        &              & $0.2925(18)$ & $0.2903(32)$ \\
\hline
        & $0.151605$  &        &              & $0.3663(14)$ & $0.3070(25)$ \\
$6.1$   & $0.152500$  & W      & $12-24$      & $0.3104(16)$ & $0.2967(33)$ \\
        & $0.153313$  &        &              & $0.2528(19)$ & $0.2862(47)$ \\
\hline
        & $0.150600$  &        &              & $0.3139(9)$  & $0.3035(19)$ \\
$6.2$   & $0.151300$  & W      & $17-31$      & $0.2669(10)$ & $0.2915(24)$ \\
        & $0.151963$  &        &              & $0.2164(12)$ & $0.2769(34)$ \\
\hline
        & $0.149259$  &        &              & $0.2977(10)$ & $0.3024(23)$ \\
$6.3$   & $0.149978$  & W      & $24-40$      & $0.2490(12)$ & $0.2874(33)$ \\
        & $0.150604$  &        &              & $0.2007(15)$ & $0.2703(56)$ \\
\hline\hline
        & $0.1334$  &        &              & $0.3988(11)$ & $0.2701(15)$ \\
$6.0$   & $0.1338$  & C      & $11-21$      & $0.3527(11)$ & $0.2658(18)$ \\
        & $0.1342$  &        &              & $0.3005(13)$ & $0.2597(27)$ \\
\hline
        & $0.1340$  &        &              & $0.3497(10)$ & $0.2729(16)$ \\
$6.1$   & $0.1345$  & C      & $12-24$      & $0.2919(12)$ & $0.2671(21)$ \\
        & $0.1350$  &        &              & $0.2223(14)$ & $0.2612(40)$ \\
\hline
        & $0.1346$  &        &              & $0.2796(7)$ & $0.2624(16)$ \\
$6.2$   & $0.1349$  & C      & $17-31$      & $0.2428(8)$ & $0.2569(20)$ \\
        & $0.1352$  &        &              & $0.2009(9)$ & $0.2519(27)$ \\
\hline
        & $0.1346$  &        &              & $0.2637(10)$ & $0.2677(26)$ \\
$6.3$   & $0.1349$  & C      & $24-40$      & $0.2281(11)$ & $0.2617(31)$ \\
        & $0.1352$  &        &              & $0.1876(13)$ & $0.2552(40)$ \\
\hline
\end{tabular}
\caption{ Our results for the two actions, Wilson (W) and non-perturbatively improved (C),
that we have used. We also give the fit interval in $x_0/a$ corresponding to a distance of
1 fm. }
\label{tab:allres}
\end{center}
\end{table}

\begin{table}[htbp]
\begin{center}
\leavevmode
\begin{tabular}[]{|c|c|c|c|c|c|c|}
\hline
$\beta$ & $a$ (fm) & $T/a$ & $L/a$ & $Z_{12}$ & $\langle x\rangle$ &
$N$   \\
\hline\hline
$6.0$   & $0.093$  & $32$  & $16$   & $1.033$      & $0.2727(46)$       & $600$ \\
$6.1$   & $0.079$  & $36$  & $16$   & $1.004$      & $0.2683(67)$       & $573$ \\
$6.2$   & $0.068$  & $48$  & $24$   & $0.985$      & $0.2549(49)$       & $416$ \\
$6.3$   & $0.059$  & $64$  & $24$   & $0.977$      & $0.2474(76)$       & $391$ \\
\hline
$6.0$   & $0.093$  & $32$  & $16$   & $1.2196$     & $0.2471(44)$       & $600$ \\
$6.1$   & $0.079$  & $36$  & $16$   & $1.1993$     & $0.2535(48)$       & $800$ \\
$6.2$   & $0.068$  & $48$  & $24$   & $1.1832$     & $0.2405(41)$       & $500$ \\
$6.3$   & $0.059$  & $64$  & $24$   & $1.1713$     & $0.2427(61)$       & $319$ \\
\hline
\end{tabular}
\caption{ Our results for the matrix element in the chiral limit at various
values of the lattice spacing. We also give the renormalization factor $Z_{12}$
as obtained from interpolating the simulation data shown in
fig.~\ref{fig:zeta}.  
The upper set of data belong to simulations with the Wilson action, the lower
set to simulations with the clover action.  In the last column we give the
number $N$ of configurations as used for evaluating $\langle x\rangle$. }
\label{tab:mainres}
\end{center}
\end{table}


\vskip 0.5cm
\noindent ACKNOWLEDGEMENTS

\noindent We have considerably profited from the experience accumulated by
the ALPHA collaboration with the Schr\"odinger functional with fermions. We
thank F.~Palombi and A.~Shindler for useful discussions.

\newpage

\def\NPB #1 #2 #3 {Nucl.~Phys.~{\bf#1} (#2)\ #3}
\def\NPBproc #1 #2 #3 {Nucl.~Phys.~B (Proc. Suppl.) {\bf#1} (#2)\ #3}
\def\PRD #1 #2 #3 {Phys.~Rev.~{\bf#1} (#2)\ #3}
\def\PLB #1 #2 #3 {Phys.~Lett.~{\bf#1} (#2)\ #3}
\def\PRL #1 #2 #3 {Phys.~Rev.~Lett.~{\bf#1} (#2)\ #3}
\def\PR  #1 #2 #3 {Phys.~Rep.~{\bf#1} (#2)\ #3}

\def\etal{{\it et al.}}
\def\ibid{{\it ibid}.}

\end{document}